*Horváth István*[1]

# A GAMMACSILLAGÁSZAT KEZDETEI

**Enlish title: The beginning of gamma astronomy**

Abstract: In the first part of this paper we summarize the discovery and the early study of the gamma-ray bursts. The second part studies the spatial and sky distribution of the bursters. The brightness distribution is also studied. In the conclusion we discuss that these are supporting the cosmological origin of the gamma-ray bursts.

Keywords: astronomy, gamma astronomy, gamma-ray bursts, satellites

Kulcsszavak: *csillagászat; gammacsillagászat; gammakitörések; kutató műholdak*

## 1. A gammakitörések felfedezése

Hosszas tárgyalások után, 1963. augusztus 5-én a Szovjetunió, az Amerikai Egyesült Államok és az Egyesült Királyság képviselői Moszkvában aláírták a Nemzetközi Atomcsend Egyezményt (Nuclear Test Ban Treaty), amely megtiltotta az atombomba-kísérleteket a légkörben, a világűrben és a víz alatt[2]. Az egyezményhez azóta több mint száz ország csatlakozott, viszont Kína és Franciaország máig sem.

Az egyezmény betartásának ellenőrzésére hozta létre az Egyesült Államok a VELA projektet, és lőtték fel a világűrbe a VELA műholdakat. A VELA szó spanyol eredetű (velar), jelentése őr, vagy felügyel. Mivel ezek a műholdak fedezték fel a gammafelvillanásokat, ezért a projektről részletesen kell szólnunk.

A programot 1959-ben indították el, viszonylag csekély költségvetéssel. Összesen tizenkét műholdat indítottak, hatot a Vela Hotel, hatot az Advanced Vela felépítéssel [Klebesadel, R. W., 2012]. A hat Vela Hotel műhold űrbéli nukleáris robbanások után kutatott, míg az Advanced Vela hagyományos robbanásokat is megfigyelt. A műholdakat a TRW (Thompson Ramo Wooldridge) cég gyártotta.

A hidegháború javában dúlt, az amerikaiak biztosra akartak menni, így a legvadabb elképzeléseket is ellenőrizni akarták. Ma már talán hihetetlennek tűnik, de attól is tartottak, hogy a Szovjetunió a Hold túlsó oldalán robbant kísérleti atomtölteteket [Vedrenne, G. and Atteia, J.-L., 2009] (2000-ben nyilvánosságra hozták, hogy maga az amerikai légierő is tervezett 1958-ban nukleáris robbantást a Holdon). Ezek megfigyeléséhez több magas pályán keringő műholdra volt szükség. Végül a műholdakat a Van Allen sugárzási övezet fölé, négyszeres geoszinkron magasságú (120 ezer kilométer magas) pályákra állították [Klebesadel, R. W., 2012].

A műholdakat párosával indították, az első Vela Hotel-párt (Vela-1A és Vela-1B) 1963 augusztusában. A második párt (Vela-2A és Vela-2B) 1964-ben, a harmadikat (Vela-3A és Vela-3B) 1965-ben állították pályára, tervezett élettartamuk 6 hónap volt, aminek végül több mint tízszeresét szolgálták ki. A műholdakat 12 külső röntgendetektorral és 18 belső

---
[1] NKE HHK KLI Természettudományi Tanszék
[2] Az egyezmény szövege a következő helyen található meg az interneten: http://www.ctbto.org/fileadmin/content/treaty/treatytext.tt.html

neutron- és gammasugár-detektorral szerelték fel [Klebesadel, R. W., Strong, I. B. and Olson, R. A., 1973].

A légköri atom- vagy hidrogénbomba-robbanás a másodperc ezred részéig tartó gammavillanást produkál, amit a kialakuló tűzgolyó fénysugárzása követ. A műholdak milliszekundumos skálán figyelték a jelenségeket, így több műhold együttes megfigyelése esetén a forrás térbeli helyzete, egyszerű háromszögeléses módszerrel, nagyjából 5000 kilométeres pontossággal meghatározható volt a Föld felszínén. Távoli források esetén ez nyilvánvalóan csak iránybeli lokalizációt jelent, amely azonban elegendő a lunáris vagy szoláris források azonosításához. Az irányszög meghatározásának hibája nagyjából 8-15 fok volt.

Az Advanced Vela párokat 1967-ben (Vela-4A és Vela-4B), 1969-ben (Vela-5A és Vela-5B) és 1970-ben (Vela-6A és Vela-6B) indították, de már nem az Atlas Agena rakétákkal, mint az előző párokat, hanem a sokkal erősebb Titan IIIc rakétával, így a tömegük is nagyobb lehetett. Tervezett élettartamuk másfél év volt, de ezek a műszerek is tízszeresen felülmúlták az előzetes terveket. Az utolsó 1984-ig működött. Az Advanced Vela műholdak hat darab cézium-jodid gammadetektorral voltak felszerelve, melyek teljes térfogata kb. 60 $cm^3$ volt, és 150-750 keV (kiloelektronvolt) közötti energiájú fotonokat voltak képesek észlelni.

A műholdak a megfigyelt adatokat továbbították a Földre. Nukleáris bomba robbanását röntgensugárzás jelezte volna, melyet a gamma- és neutrondetektorok megfigyelése erősített volna meg. A Hold túloldalán történt robbanást közvetlenül nem észlelték volna a Vela műholdak, de a felvert nukleáris por a robbanás erejétől olyan gyorsan tágult volna, melyet a műholdak a robbanás során aktivált atommagok gammasugárzását megfigyelve tudtak volna azonosítani.

Ray Klebesadel a Los Alamos Scientific Laboratory (ma LANL, Los Alamos National Laboratory) munkatársa (aki a Vela műholdak tervezésében és építésében is részt vett) elemezte a megfigyelt adatokat. Azokat a megfigyelési eredményeket is gondosan megőrizték, melyek biztosan nem nukleáris robbanást jeleztek. 1972-ben Ian Strongot kérték meg, hogy Klebesadellel és Roy Olsennel közösen értékeljék ki ezeket az adatokat, akik 16 olyan eseményt találtak, melyek bizonyosan nem földi, szoláris vagy lunáris eredetűek voltak (1969 júliusa és 1972 júliusa közötti jelenségekről volt szó). Gammatartományban olyan jelentős volt az emisszió, hogy ki lehetett zárni, hogy egy röntgenforrás nagyenergiás részéről legyen szó. Ebből az eredményből született meg az első gammakitörés (gamma-ray burst vagy röviden GRB) publikáció [Klebesadel, R. W., Strong, I. B. and Olson, R. A., 1973].

Cline és Desai az IMP-6 műhold fedélzetén lévő röntgendetektorral a napflerek megfigyelését végezték. Értesülve a felfedezésről, az elsők voltak, akik megerősítették a gammavillanások létezését [Cline, T. L., Desai, U. D., Klebesadel, R. W. and Strong, I. B., 1973]. Ezektől függetlenül az OSO-7 műhold gammadetektorai is megerősítették ezt a felfedezést. [Wheaton, W. A., et al., 1973].

## 2. Bolygóközi hálózat

A gammakitörések véletlen felfedezésének hatására az azt követő években több műholdra és műbolygóra gammasugárzást megfigyelni képes műszereket szereltek fel. Az orosz mesterséges égitestekre általában felszereltek gammadetektorokat. Ezek esetében azonban az adatokhoz való hozzájutás okozott problémát. Volt gammadetektor az 1977-ben fellőtt Prognoz 6 és az 1978 októberében felbocsátott Prognoz 7 fedélzetén is. Mindkettő a Föld körül keringett. NaI és CsI gammadetektorok voltak a KONUS [Mazets, E. P., et al., 1981], az ISEE-3 (International Sun-Earth Explorer) [Anderson, K., et al., 1978] és az SMM (Solar

Maximum Mission) [Forest, D., et al., 1980] műbolygókon. Gammadetektorok voltak még a High Energy Astronomical Observatory (HEAO), az International Cometary Explorer (ICE), a Helios 2, a Hakucho, a Hinotorti, a Ginga, a Phobos [Higdon, J. C. and Lingenfelter, R. E., 1990] és az Apollo-16 fedélzetén is [Metzger, A. E., et al., 1974].

A gammavillanások tisztázatlan eredete miatt fontos lett volna a pontos lokalizáció az esetleges források tanulmányozására. A helymeghatározás egy detektor esetén lehetetlen, hiszen így csak a fotonok beérkezését lehet rögzíteni. Megfelelő elektronikával ez a beérkezés milliszekundumos pontossággal volt rögzíthető. Két egymástól távoli detektor megfigyelése esetén a megfigyelt jelek időkülönbségéből az égi helyzet meghatározható. Pontosabban szólva két független megfigyelésből a számolható pozíció egy vékony körgyűrű az égen. Pontos lokalizáláshoz minimum három egymástól távoli műszer egyidejű megfigyelése szükséges [Atteia, J-L., et al., 1987]. Itt a hangsúly az egymástól távoli kifejezésen van. Időben fél-egy perc, esetleg több perces eltérés szükséges. Vagyis a megfigyelőhálózat egy eleme keringhet a Föld körül, de a többi már nem. Így legalább két Földtől távoli műszer szükséges. A német Helios 2 műbolygó a Nap körül keringett, fedélzetén gammadetektorral, így kiválóan megfelelt e kutatási célra.

A harmadik tájékozódási pont a Vénusz körül keringő három műhold volt: az orosz Venyera 11 és 12 [Barat, C., et al., 1981] valamint az amerikai PVO (Pioneer Venus Orbiter) [Klebesadel, R. W., et al., 1980]. Mindhármat szintén 1978-ban bocsátották útjára. Az ún. első bolygóközi hálózat 1980-ig működött[3]. Megfigyelték a gammakitöréseket, majd a már említett háromszögelési módszerrel meghatározták a források irányát (84 kitörésre állapítottak meg égi koordinátákat [Hartmann, D. and Epstein, R., 1989]). De a kibocsátás helyén még a legnagyobb távcsövekkel sem találtak semmit.

1990-ben felbocsátották az Ulysses műbolygót, majd 1991-ben a Compton Gamma Űrtávcsövet (CGRO, Compton Gamma Ray Observatory) [Fishman, G. J., et al., 1994], így ismét létrejött egy bolygóközi hálózat [Cline, T. L., et al., 1991].

### 3. Milyen távol vannak?

A 80-as évek végéig közel 500 gammafelvillanást észleltek a különböző műholdak [Fishman, G. J. and Meegan, C. A., 2012]. Ezek nagy részét a PVO [Evans W. D., Fenimore E. E., Klebesadel R. W., Laros J. G. and Terrel N. J., 1981] és a KONUS [Mazets, E. P., et al., 1981] figyelte meg, de a felvillanásoknak az eredete még tisztázásra várt. Nem volt ismeretes egy azonosított forrásuk sem, de még azt sem tudtuk, hogy milyen messze lehetnek a források. Tehát az sem volt ismeretes, hogy a forrás milyen erős, ugyanis a mért gammasugárzást okozhatta egy közeli gyenge forrás, vagy egy távoli, de ennél sokkal erősebb forrás is.

Az egyik elfogadható feltételezés a galaktikus neutroncsillagokból jövő sugárzás volt [Higdon, J. C. and Lingenfelter, R. E., 1990], [Harding, A. K., 1994], [Colgate, S. A. and Li, H., 1996]. Ez esetben a források a galaxis síkjában vagy a galaktikus haloban helyezkednének el, kiloparszek illetve 10-20 kiloparszek távolságban.

A különféle asztrofizikai objektumok távolságának a meghatározásában nagy szerepe van az égbolton való eloszlásuknak, elhelyezkedésüknek. Lehetnek a források a Naprendszerben vagy ahhoz közel. Ha például a források a bolygók felszínén vannak, akkor az égen ábrázolva az ekliptika mentén helyezkednének el, korrelálva természetesen a bolygópozíciókkal.

---

[3] http://www.powerset.com/explore/semhtml/InterPlanetary_Network?query=grb

Ha valamely asztrofizikai megfigyelés esetén közeli csillagok lennének a források, akkor a helyük az égre vetítve véletlenszerűen helyezkedne el, tehát betöltenék a teljes eget. Ha egy adott távolságnál közelebb lévő források számát N-nel jelöljük, akkor a kétszer akkora távolságnál közelebb lévő források száma körülbelül 8N lenne, ha a források eloszlása homogén, hiszen a megfigyelt objektumok száma arányos a térfogattal.

Ha a galaxisunkban levő forrásokat figyelnénk a Földről, akkor a kétszer olyan távolságban lévő források száma csak 4N lenne, hiszen a források egy lapos korongban helyezkednek el (pl. az 1. ábra bal alsó ábrája), és a kétszer akkora sugarú kör területe csak négyszer nagyobb.

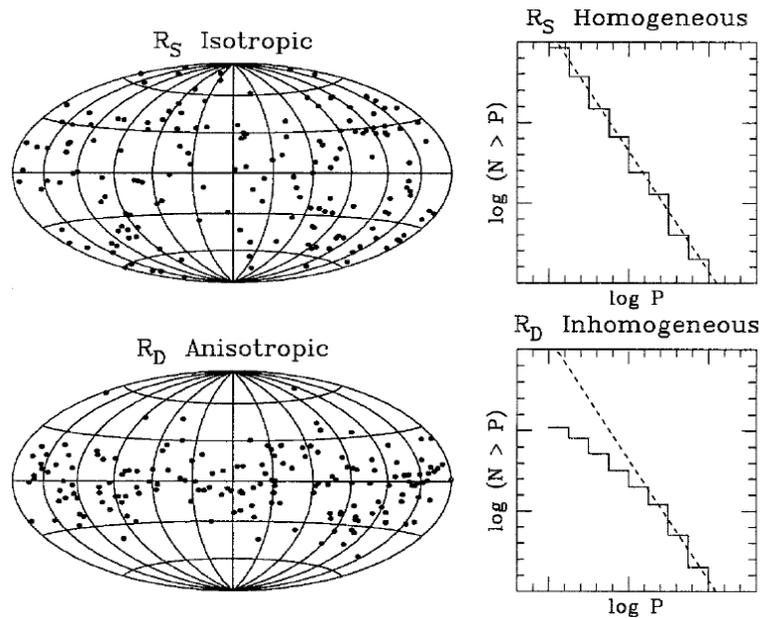

*1. ábra A Gamma-ray Bursts kötetben [Fishman, G. J. and Meegan, C. A., 2012] bemutatott ábra szemlélteti, hogy homogén térbeli eloszlás esetén (jobb felső ábra) statikus euklideszi térben az égboteloszlás egyenletes (izotróp) kell legyen. Amennyiben a térbeli eloszlás nem homogén (jobb alsó ábra), úgy az égboteloszlás sem az.*

Tehát a források számának a fényességüktől való függése információt ad a források térbeli elhelyezkedésével kapcsolatban. Nézzünk egy egyszerű példát. Töltsék ki a források a teljes teret egyenletesen, és legyen minden forrás egyforma fényességű. Ez esetben a legfényesebbnek látszó forrás van hozzánk legközelebb. A négyszer halványabb források kétszer messzebb vannak, de mivel a kétszer nagyobb sugarú gömb térfogata nyolcszor nagyobb, ezért a négyszer halványabb források száma átlagosan nyolcszor több. A kitevőben lévő kettes és hármas eredményeképpen a logaritmikus ábrázolásnál a jelenséget egy mínusz háromketted meredekségű egyenes jól közelíti. Ez a híres „-3/2"-es törvény.

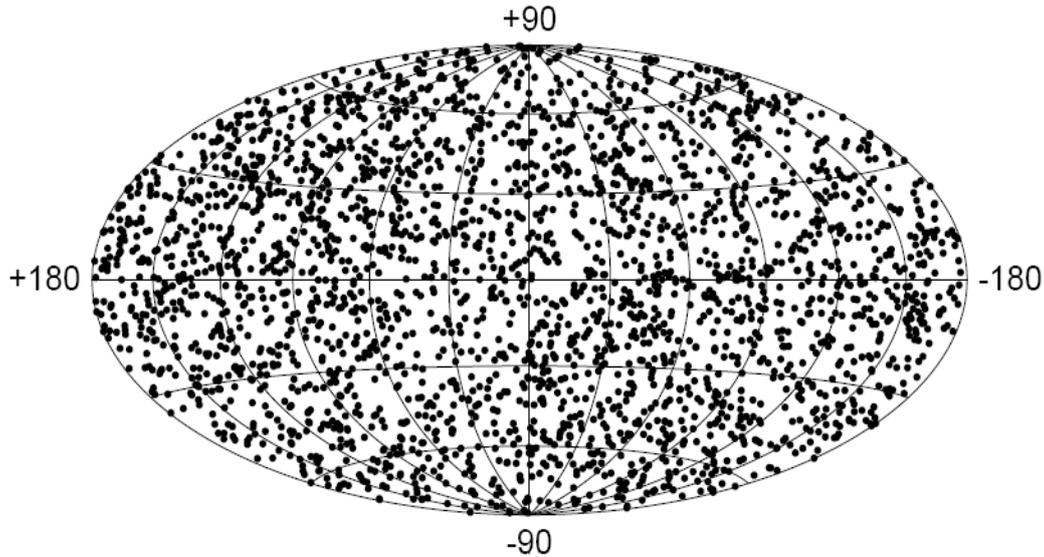

*2. ábra A gammakitörések eloszlása az égbolton a Compton Gamma Obszervatórium Burst And Transient Source Experiment (BATSE) adatai alapján.*

A Compton Gamma Ray Observatory (CGRO) műhold, egyik nagy tudományos eredménye volt a kitörések égi eloszlásának megmérése. Ez az eloszlás egyenletes eloszlást mutatott az égbolton (lásd a 2. ábrát) [Meegan, C. A., et al., 1996], [Vavrek, R., et al., 2008]. Ez a megfigyelés kizárta a galaktikus eredetet. Ugyanis ha a források nagy része a teljes galaxisban található, akkor a galaxis síkjának látszódnia kellene az égi eloszláson. Ennek ellenére egészen 1996-ig tartotta magát a kiterjesztett galaktikus halo-eredet hipotézise (extended galactic halo) [Brainerd, J., 1992], [Podsiadlowski, P., Rees, M. J. and Ruderman, M., 1995].

Ha a források egyenletesen oszlanak el az égen, akkor csak három térrész képzelhető el a források eredetére:

1, A Naprendszer-közeli tér.
2, A fényes csillagokhoz hasonlóan egy néhány tucat, maximum néhány száz parszeknyi térrész.
3, Száz megaparszek vagy annál lényegesen nagyobb sugarú tér.

Az elsővel kapcsolatban dolgozták ki az üstökös felhő elméletet, amely a 100-1000 CsE (csillagászati egység = Nap Föld távolság) távolságban keringő üstökösöket tekintette forrásnak [Bickert, K. F. and Greiner, J., 1993], [White, R. S., 1993], [Luchkov, B. I., 1994].

A háromféle eredet között segít választani a fényességeloszlás-ábra, az ún. logN – logS diagram [Cohen, E. and Piran, T., 1995], [Fenimore, E. E. and Bloom, J. S., 1995]. A 3. ábra mutatja közel ezer kitörés látszó fényesség szerinti kumulatív eloszlását [Horváth, I., Mészáros, P. and Mészáros, A., 1996]. A függvény értéke azt adja meg, hogy hány kitörést figyeltünk meg adott idő alatt, amely az adott fényességnél (S) fényesebb volt.

Euklideszi tér és homogén térbeli eloszlás esetén a már az előzőekben leírt okok miatt egy körülbelül -1,5 meredekségű egyenest (szaggatott vonal a 3. ábrán) kell kapnunk a log-log ábrán. Ez csak a fényes kitörésekre igaz a gammakitörések esetén [Fishman, G. J., et al., 1994], [Veres, P., et al., 2010], [Perez-Ramirez, D., et al., 2010]. A halvány kitörések hiányát persze lehetne magyarázni elnyeléssel, de csillagászati megfigyelésekből tudjuk, hogy sem a Naprendszer közelében, sem a néhány száz parszekes környezetünkben nincs lényeges elnyelő anyag. Egy mindent áteresztő, csak gammában elnyelő közeg feltételezése pedig abszurd lenne.

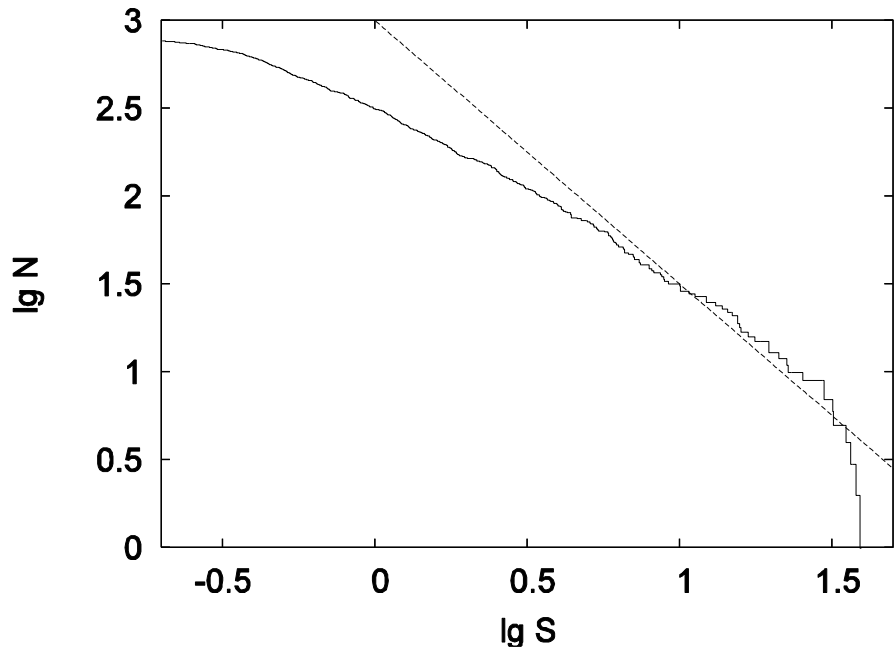

*3. ábra A gammakitörések megfigyelt fényességeloszlása.*

     Mindezek alapján nyilvánvalóan csak a harmadik eset lehetséges. Ugyanis valóban euklideszi tér esetén mindig igaz a mínusz másfeles törvény. Tudjuk viszont, hogy a világunk nem euklideszi, hanem riemanni [Riemann, B., 1854], [Einstein, A., 1916], [Friedmann, A., 1922], [Paál, G., Horváth, I. and Lukács, B., 1992]. A kozmológiai megoldások ([Walker, A. G., 1936], [Peebles, P. J. E., 1993]) valóban megmagyarázhatják a mért logN – logS eloszlást [Cohen, E. and Piran, T., 1995], [Fenimore, E. E. and Bloom, J. S., 1995], [Horváth, I., Mészáros, P. and Mészáros, A., 1996]. Azonban ehhez a kitörések forrásait a legtávolabbi kvazárok távolságáig kell elképzelnünk. Ezt az elképzelést már a hetvenes években felvetették [Usov, V. V. and Chibisov, G. V., 1975], [Prilutskii, O. F. and Usov, V. V., 1975], de csak Paczynski cikke [Paczynski, B., 1986] után vették többen komolyabban. Pedig az 1. ábráról is világosan látszik, hogy Euklideszi tér esetén, egyenletes eloszlásnál a bal felső égi és jobb felső térbeli eloszlást, vagy a két alsó eloszlást figyelhetjük meg. De a gammakitörések a bal felső és jobb alsó eloszlásokat követik. Ez csak akkor lehetséges, ha az általuk kitöltött tér nem euklideszi. A nagyon távoli (kozmológiai távolságban levő) objektumok által kitöltött tér nem euklideszi, mert a négydimenziós téridő görbült.

     Mindezek ellenére többeknek ez annyira hihetetlennek tűnt, hogy a tudóstársadalom legalább fele inkább az egyre extrémebb kiterjesztett halo elméletek kidolgozásán fáradozott. Tették mindezt egészen 1997-ig, az első gammakitörés-utófény, illetve -forrás azonosításáig, melyet a BeppoSAX műhold tett meg.